\documentstyle[12pt,epsfig]{article}
\newcommand{\be}{\begin{eqnarray}}
\newcommand{\ee}{\end{eqnarray}}

\begin{document}
\title{ SUMMARY OF THE WORKSHOP \\ ``THE QCD PHASE TRANSITIONS'' \\
Brookhaven National Laboratory, \\ November 4-7 1998 }
\author{Edward Shuryak\\
 Department of Physics and Astronomy \\
 State University of New York, Stony Brook Ny 11794, USA}
\maketitle
\section{Overview}
The title of the workshop, ``The QCD Phase Transitions'', in fact
happened to be too narrow for its real contents. It would be more
accurate to say that it was devoted to 
different phases of QCD and
QCD-related
gauge theories, with strong emphasis on discussion of the underlying
non-perturbative mechanisms
which manifest themselves as all those phases. 

  Before we go to specifics, let us emphasize one important aspect 
of the present status of non-perturbative Quantum Field Theory in general.
It remains true that its studies do not get attention proportional
to the intellectual challenge they deserve, and that the
theorists working on it remain very fragmented.
The 
efforts to create Theory of Everything including Quantum
Gravity have  attracted the lion share of attention and young talent.
Nevertheless, 
in the last few years there was also a tremendous progress and even some shift
of attention toward emphasis on the unity of non-perturbative
phenomena.  For example, we have seen some efforts 
to connect the lessons from recent progress in Supersymmetric theories
with that in QCD, as derived from phenomenology and lattice. Another
example is Maldacena conjecture and related development, which
connect three things together, 
string theory, super-gravity and the (N=4) supersymmetric gauge theory.
Although
 the progress mentioned is remarkable by itself,  if we would listen 
to each other more we may have chance to strengthen the field and 
reach better understanding 
of the spectacular non-perturbative physics.

  That is why 
the workshop  was an attempt to bring together
people which normally belong to different communities 
and even cultures (they
use different tools, from lattice simulations to models to exact solutions),
in order to discuss common physics. It was a very successful, eye-opening 
 meeting for many participants, as some of them said in the last round of
discussions.
Even organizers (who of course have  contacted
many speakers in advance)
were amazed  by completely unexpected things 
 which were popping out of one talk after 
another. Extensive afternoon discussion, in which we always return back
to the morning talks, has helped to clarify many issues.

For QCD one of the main source of ``input" remains experimental data
about hadrons. The second, now nearly as important as the first, is
  provided
by numerical lattice simulations. Those can also consider various
flavor contents, change the quark masses,
easily access finite temperatures (finite density remains so far a 
problem).  Furthermore, they can study observables not in average, but
on
configuration-by-configuration basis, and reveal more details about a dynamics.
 The third major  input is provided by exactly solvable (or partially
 solvable)     
models, mostly the Super-symmetric (SUSY) ones.

  Let me on the onset indicate some similarity between various approaches
discussed on the workshop. Many (if not most) of the talks  in this way or another  separate ``quantum noise" (the perturbative
phenomena) from ``smooth" or even classical fields, related to 
 non-perturbative dynamics. The tools used for this
general aim are however very
different: (i) Blocking lattice configurations, or ``cooling" them;
 (ii) Considering super-symmetric
theories in which many diagrams cancel; (iii) Considering
large $N_c$ limit, in which there should be some ``master field" dominating
the path integrals (Mattis again); (iv) going to complex-valued configurations,
which are some non-trivial saddle points (Velkovsky).

  But whatever the tools, the classical configurations themselves revealed
in those analysis happened  to be nothing else but our old friend, the
$instanton$. Their ensemble
  saturates the topological susceptibility,  solving the
U(1) problem\footnote{ Not ``in principle" (which 't Hooft did back in
  1976), but for real,  quantitatively reproduces the value needed to
explain correct $\eta'$ mass.}.
They  also do saturate the lowest Dirac eigenmodes, explaining
chiral symmetry breaking (again quantitatively, producing
 accurate value for the quark condensate) and even hadronic
 correlators,
see recent review \cite{SS_98}.
I will argue below that instantons explain also the origin of the famous
 ``chiral scale" 1 GeV in QCD \cite{RRS}.
Furthermore, recently instantons emerged as the main driving force
in Color Superconductivity. 

Instantons also provide few
exact results for SUSY theories. They reproduce expansion
of the Seiberg-Witten ``elliptic curve" for N=2 SUSY QCD \cite{susy_inst}, and also
provide the ``master field'' of the N=4 theory \cite{Mattis_etal}, as discussed here by Mattis.

  However many properties of the instanton ensemble are far from being
  clear.
The major example 
 (discussed 
especially by de Forcrand) is complicated behavior near the critical
temperature $T_c$: qualitative changes in their ensemble are obvious but
the structure above $T_c$  is not
yet understood.

  The only exceptional non-perturbative
phenomenon which instantons do $not$ explain is
 confinement \cite{CBNS,DP_conf}: 
this issue was discussed by Negele. 

\section{High density QCD}
The field of high density QCD was mostly dormant since late-70's-early 80's,
when implications of perturbative QCD for this case was worked out.
However realization last year (simultaneously by ``Stony Brook'' and ``Princeton''
 groups \cite{RSSV,ARW} ) that instantons can induced
not only strong pairing of quarks with anti-quark in vacuum and break
chiral symmetry, but also 
a quark-quark pairing at high density, has created a splash of activity. 
Such Color
Super-Conducting (CSC) 
 phase was under very intense discussion at the workshop.

 It was introduced in the first review talk  by F.Wilczek
(Princeton), who emphasized the so called color-flavor locking phase \cite{ARW2}
which appears for
three massless quarks ($N_f=3$). Discussion of its rather unusual qualitative
features was
continued by T.Schafer (Princeton), who has presented
some  quantitative results \cite{RSSV2} 
following from account for instanton interaction. One important
result was a  demonstration
that,  as one increase the
mass
of the strange quark and goes back to the $N_f=2$ theory, 
no phase transitions actually happens and interpolation between two
different structures of CSC
is in fact continuous. Another interesting issue, for $N_f=3$ case,  is 
whether there can in principle be a continuous transition from
hadronic to CSC phase. Schaefer and Wilczek
\cite{SW} suggested that the answer is positive.

G.Carter (Copenhagen) had further
discussed the $N_f=2$ case in the instanton model in some details \cite{CD},
including correct instanton-induced form-factors. R.Rapp (Stony Brook)
have provided another view on this subject \cite{RSSV2}, using statistical rather
than mean field description of the instanton ensemble, and
 discussing the role of
instanton-anti-instanton
molecules in this transition.

After the workshop an interesting paper written by Son \cite{Son} have
shown
that in the high density (weak coupling) limit (when the instantons
are Debye-screened) the leading behavior is not provided by electric
(Coulomb) part of the one-gluon exchange, but by a magnetic one.  

The talks have so many details that I would not go into it.
In summary, QCD demonstrate a kind of ``triality". There are
three  major phases of QCD: (i) hadronic, dominated by $\bar q q$ attraction
leading to chiral symmetry breaking; (ii) CSC at high density,
dominated by qq attraction and condensation, and
(iii) QGP at high T, in which there are no condensates but instantons
and anti-instantons themselves are bound by a fermion-induced
forces. 

 A complementary approach to high density QCD, now
based  on random matrix model, was reviewed by
M.Stephanov (Stony Brook). He outlined what exactly goes wrong in
``quenched'' QCD at finite density, and also how the correct behavior
of the Dirac eigenvalue at increasing $\mu$ should look like: the resulting
picture resembles ``a dividing chromosome'', rather than a ``cloud''  
coming from quenched theory. He also pointed out the existence of the
tri-critical point at the phase diagram of the random matrix model
\cite{JSV}, as well as 
importance and even possible ways to search
for it in heavy ion collisions \cite{SRS}.

  Various ideas of how one can proceed to study the high density on
the lattice were also  discussed.
 At the end of the talk F.Karsch described new approach,
with
finite baryon density (instead of chemical potential).
M.Alford (MIT)  has described possible analytic continuation to complex
chemical
potential.

Finally
M-P.Lombardo (Gran Sasso) had presented very interesting data for
2-color QCD. In this theory the determinant is real even with chemical
potential, and so the usual lattice calculations are possible.
The results are consistent with CSC phase being developed.

\section{High temperature QCD}
  Lattice results on finite temperature transitions were reviewed  by F.Karsch
(Bielefeld) and also by C.DeTar (U. of Utah) 
. Excellent data for pure gauge theories exist by now, and they show
transition at $T_c\approx 260 MeV$.
The ratio to the string tension  $T_c/\sigma^{1/2}$
is close to $(3/(d-2)\pi)^{1/2}$ as
predicted by the string model of deconfinement.  M.Wingate
(RIKEN/BNL) has presented new data for deconfinement in 4-color gauge
theory, which also support this trend.

  However, 
  as it is well known by now, QCD with  light quarks show
{\it much smaller} critical temperature $T_c$.
This suggests that it has nothing to do with deconfinement, as it is
described by the string model.

 For 2 light quarks ($N_f=2$)
 $T_c\approx 150 MeV$ and is driven by chiral symmetry
restoration. The order  of the transition in the 
$N_f=2$ theory is second, as expected, but ``current analysis did not
reproduced
the expected critical behavior for a system in the universality class
of O(4)-symmetric spin models'', Karsch concluded. The situation 
remains to be quite confusing,
the current set of indices  do not fit into any of the established 
universality classes. Maybe the issue is complicated by ``approximate
restoration
of the U(1) symmetry'' \cite{Shu_94} which add 4 more light (although still
massive)
modes. If so, the transition may be driven  to weak first order instead.
 DeTar have also shown how lattice artifacts present for $N_t=4$ 
and creating doubts about relevance of this case for continuous
limit, are actually dissolves for larger values
of $N_t$ (up to 12) studied.

 DeTar also mentioned
interesting simulations by Kogut et al \cite{kogut_etal} 
who found weak first order in a simulation in which on top of standard
lattice action a small 4-fermion term was added. 
Let me comment on it: Kogut et al have considered this interaction as a pure
methodical tool, they did not specified or speculated about its possible
structure.
I have however made a point that in fact there is the natural reason
why such small interaction should exist:  there are small-size
($\rho\sim a$)  instantons  which ``fall through the lattice''. Their contribution
should therefore be explicitly added, as another operator into the
lattice action.

For the $N_f=4$ theory, discussed by Mawhinney, the condensate is so small that
the critical temperature is not even measured yet. It
however  supports a prediction
of the instanton liquid model \cite{SS_98} that instanton-induced chiral
symmetry breaking should be small at $N_f=4$ and gone by $N_f=5$, even at $T=0$.

 The central part of the talk by
  R.Mawhinney (Columbia) was first results on chiral restoration
  phase transition 
using new
``domain wall" lattice fermions \cite{Maw}. The first result 
 is that in this case the chiral symmetry is very accurate\footnote{It is
broken only by an exponentially small tails of the fermionic wave functions,
bound to ``plus" and ``minus" walls.}, and so  one can clearly recognize some zero modes of
instantons.

   In particularly, he discussed also an old question: {\it what
     happens in
the quenched (pure gauge)
theory above $T_c$?}.
 Without a determinant, there is no reason for the instantons
to be strongly correlated, and if they are more or less random
the chiral symmetry should $not$ be restored. That contradicted to 
earlier
lattice data, who concluded that chiral symmetry is restored above the
deconfinement transition.

 One well-understood issue arise here, which may affect recent (not so
 large-volume) simulations. The
total topological charge of the configuration with randomly placed
instantons
 is
Q=$|N_+-N_-|\sim \sqrt{N_++N_-}$. Therefore spectrum of the
Dirac eigenmodes of quenched configurations should have
a term 
$$ {dN\over d\lambda}= \delta(\lambda)*O(V_4^{1/2})$$
where $V_4$ is the 4-volume.
According to Banks-Casher formula $ {dN/d\lambda}(0)=\pi |<\bar q q>|/V_4$,
but this density 
 does not lead to infinite condensate because it drops out in the thermodynamical
limit. 

  New Columbia
data shown by Mawhinney are consistent with this interpretation for  
$T<T_c$, but 
 {\it above} $T_c$ the comparison for few volumes available suggested
that the coefficient was actually $O(V)$, and the contribution
to the condensate therefore is there.  
He concluded that $<\bar q q>$ is in fact $infinite$ above $T_c$, not
zero as people have claimed before. This is in sharp contrast to
earlier works: the measured condensate has changed from 0 to $\infty$!

  This result can probably be resolved as follows\footnote{This
    comment
was made in the discussion by T.Schaefer.}. At high T the overlap
matrix elements between instantons are qualitatively different: instead of
decreasing with distance as $R^{-3}$ (as at T=0), there appear exponential suppression $exp(-\pi T r) $ for spatial distance r. Therefore, the whole zone of instanton-related modes shrinks and it looks as $O(V)\delta(\lambda)$ if the quark mass is
not small compared to its width.   
  
  True shape of the the zone based on weakly overlapping instantons and 
anti-instantons\footnote{It is better to
consider the case when  their number is exactly the same, Q=0, so that there are
no exactly zero topological modes. } 
was discussed by Verbaarschot
Stony Brook). His result \cite{V} (recently also confirmed by M.Teper et
al\cite{T})
is that in
quenched QCD the eigenvalue density actually does grow indefinitely at
the origin, but as $dN/d\lambda= O(V)log \lambda$.

What this means for Columbia results is that for sufficiently small
masses (or large length in the 5-th dimension) the singularity in the 
condensate
is going to change from 1/m to log(m).  The same behavior should also
be there at low T as well, so the quenched theory always has an
infinite condensate.

I.Zahed (Stony Brook) has discussed new ideas \cite{JNPZ}
 about ``chiral disorder", connecting motion of light quarks in
the QCD vacuum to that of electrons in ``dirty metals''. He also
proposed two potentially possible regimes for chiral restoration
(i) fractal support for the chiral condensate; (ii) either some intermediate phase or 
specific places on the phase diagram where finite $<\bar q q>$
(density of eigenvalues) coexist with zero $F_\pi=0$ (no conductivity) due to eigenmodes localization.

  J.Verbaarschot (Stony Brook) have discussed a number of topics about the
 Dirac eigenvalues. The main point was that zero-momentum sector
reduces to Chiral Random Matrix Theory, but it deviates from it at larger
eigenvalues \cite{V2}  
He disagreed with Zahed on his last point, arguing 
(following Parisi) that the localized modes are independent and
therefore the fermionic determinant should be a product of the
eigenvalues. It
strongly mis-favored by any 
unquenched theory due to smallness of the fermionic determinant, and
so
he concluded localization scenario is not viable.

M.Engelhardt (Tubingen) have argued  that the deconfinement in
pure gauge theory 
can be described de to vortex percolation, rather than monopoles.

\section{Lattice instantons at zero and non-zero T}

The issue was reviewed by
J.Negele (MIT), see \cite{Negele_review}.
He shown that topological susceptibility is stabilized in many
simulations,
and the value (dominated by instantons) agrees well with
Witten-Veneziano
formula. The measurements of the size, defined by extrapolation to the
uncooled
vacuum, give $\rho=.39\pm 0.05$ fm. This number, as well as the shape
of the size distribution,  agrees well with the
phenomenology and the instanton liquid calculations.
For finite T the size decreases by about 25\% by $T=1.3 T_c$, and
shrinks at higher T, also in good agreement with the Debye screening mechanism
\cite{Shu_78,PY}.
 
 Negele has shown that most of
 the smallest fermionic zero modes
are related to instantons, both in quenched and full
simulations.
The important conclusion is  that the quark condensate
is definitely completely dominated by instantons. Furthermore, restricting
the quark propagator to 
contribution of the lowest modes only, one actually reproduces
the correlation functions, not only for such ``collective mode" as pions
but also for other channels, in particularly $\rho$.
Again, this is in agreement which we have found previously
by doing correlators in the instanton liquid models.

Another issue Negele discussed based on \cite{CBNS}
was the role of instantons in
the heavy quark potential and  confinement.
The conclusion is that the ``instanton liquid" does not confine, and
contribute to heavy quark potential at the 10-20 \% level.  
The potential found agrees well with other numerical calculations done before,
and with analytical one due to Diakonov and Petrov. 

There are however
three extra points which can be made in connection to this issue.
One is that we have found during this investigation that the potential
is sensitive to the shape of the Wilson loop, and only if its time dimension
T is much larger than spatial one L one gets a correct potential.
Diakonov and Petrov recently wrote a rather provocative paper\cite{DP_conf},
arguing that all existing lattice measurements of the confinement
at distances above 1 fm are actually from loops with $L>>T$,
and are therefore suspicious. Unfortunately, simple statistical argument
shows that it is practically impossible to go to large enough L in a 
correct way.

The second point is related with another idea, suggested by Diakonov
et al \cite{jap_D}, namely that a tail of the distribution at the  large-size
side may decrease as $dN/d\rho \sim \rho^{-3}$ and lead to infinite
confining potential. I think it cannot work, or rather
in any way explain what we know about confinement
from the lattice. One basic reason is
that it would not generate small-size strings, and also
generate long-range gluonic correlators. The other is that 
huge configuration-per-configuration fluctuations of the string tension
would be the case, again contrary to to observations.

My third comment is a phenomenological observation, which is by no means new
but I think reveal something
profoundly  important. It is found that quarkonia made of
heavy quarks (c,b) and related to confining (and Coulomb) potential
have surprisingly small interactions with light quark hadrons.
Examples are numerous, let me give one only. Compare 
two decays with the same quantum numbers of the
participants and about  the same released
energy, $\rho'\rightarrow \rho\pi\pi$ and $\psi'\rightarrow \psi\pi\pi$. 
 The ratio of widths is about a factor 1000! Where this huge factor come from?
Only from very different nature of light-quark hadrons (collective excitations
of the quark condensate, in a way, as Negele demonstrated)
 and quarkonia, bound by the confining strings. Why this interaction is
$so$ small remains unknown.

T.DeGrand and A.Hasenfratz (Boulder) have presented different aspects
of their extensive studies of lattice instantons using improved
actions
\cite{H}.

DeGrand reached  conclusions similar to Negele's
about instantons dominating  the smallest eigenvalues, but has
shown that instantons alone lead to
bad results for  the correlators, even the pion one.
The difference should be due to  different lattice fermions (KS in his work,
Wilson in Negele's): in the debate to follow I made a point 
that in KS case lattice artifacts forbid ``collectivisation" of eigenmodes
(leading  to a scenario similar to what was advocated by Zahed).   

A.Hasenfratz (Boulder)  described the current status of their work aimed
to used ``perfect lattice actions" to revealed the true soft content
of the quantum configurations. Impressive results for topological
observables such as
instanton size distribution were presented. The instanton sizes were shown to drift upward, presumably due to mutual attraction, and so
 the ``extrapolation back" seem like  a good idea.
She had also demonstrated that maybe the best way to ``hunt for
instantons''
is not via very noisy gauge fields, but from lowest fermionic eigenmodes.

One issue discussed in connection to this talks was related to
what we actually mean by `` total"
instanton density. It is clear that as it is done it depends on
particular program recognizing instantons. Closed 
$\bar I I$ pairs (or ``fluctons" as I have called them in studies of
tunneling in quantum mechanics \cite{fluctons}) can only be separated
from perturbative fluctuations by some {\it ad hoc} condition, since
there is no real difference between the two. Still, let me point out,
to a large extent such pairs  can still be well described
by semi-classical fields: only instead of the classical fields (minima  
of the action) we should look at the ``streamline" configurations.
Their shapes (and references to the previous works) can be found
in \cite{V_sl}: those can well be used for ``flucton recognition".  

In summary: the instanton-antiinstanton pairs form the famous
valley of Q=0 configurations, going smoothly to zero field one.
Its population in the vacuum may and can be studied, especially in
connection to the long-pending question about understanding
of ``non-perturbative'' aspects of high-order perturbative terms.
However, those close pairs do not provide the main object of the
instanton physics, the lowest Dirac eigenmodes, and so they
would be simply ignored by any fermionic algorithms (like the one discussed
by Hasenfratz).

Ph.de Forcrand (Zurich) 
had also described his version
of the ``improved cooling'' as a way to look for the
instantons. He has also
observed good agreement between Banks-Casher relation used for the
instanton eigenmodes, and the value of the quark condensate. 
The main topic of his talk however is related with a puzzling
question, {\em what happens at $T>T_c$} for QCD with dynamical quarks?

The proposal by Ilgenfritz and myself \cite{IS_94} was that the ensemble
of instantons is broken into so called
instanton-anti-instanton molecules. This idea has worked well in
the instanton liquid model simulations, see review \cite{SS_98}.

 However, de Forcrand et al results \cite{dF} neither disprove nor completely supported
this scenario.
On the $pro$ side, de Forcrand had demonstrated us
 that all configuration there have Q=0, and that the Dirac eigenvalue spectrum
even develops something like a forbidden gap. Many of the smallest eigenmodes
do indeed display two maxima in space-time, corresponding to instanton and
anti-instanton. There is   also some support to our prediction
that the molecules should be predominantly oriented in time direction. 
However, on the $con$ side, as seen from de Forcrand's  movie
displaying instantons at different T, 
 pure inspection of the action
does not provide any clear identification of the $\bar I I$ pairs 
or other clusters in this ensemble. Therefore a change in the spectrum
remains
a mystery.

In connection to this issue, let me recall  recent work by 
Ilgenfritz and Thurner \cite{IT}. Although for quenched configurations  
only, they have developed a way to correlate relative color orientations
of instanton and anti-instanton. They have measured distribution of
the following quantity\footnote{In fact in order it to be non=zero, it is also necessary to flip sign of the electric component in one of the fields.}
$$ ``cos\theta"= {<G_{mu\nu}(z_I)U G_{mu\nu}(z_{\bar I})U^+> \over
| G_{mu\nu}(z_I)|| G_{mu\nu}(z_{\bar I})|} $$
where U is transport between  centers $z_I,z_{\bar I}. $
The surprising result is that the distribution is very different at
low T and  $T>T_c$: the former correspond to random distribution, with
cos$\theta$ peaked around 0, while in the latter case it is peaked at
1 and -1. It probably means, that {\em even in quenched theory without the
determinant} there is some formation of the ``molecules". 

Let me summarize the somewhat puzzling situation
once again: de Forcrand et al have found only marginal
support for the molecular scenario in $full$ theory (where it was predicted),
while Ilgenfritz and Thurner seem to find them in $ quenched$ theory (where we
did not expected to find them). New simulations, with smaller quark
masses (or better, with domain wall fermions) and new way of analysis
are needed to clarify it.

\section{QCD at larger number of flavors}

This is one more direction of the QCD phase diagram, in which
we expect chiral symmetry restoration. 
As it is well known, right below the line at which asymptotic freedom
disappears ($N_f=11*N_c/2$) the new phase must be a conformal theory because
the beta function crosses zero and therefore the theory
has an infrared fixed point. We do not however know till what $N_f$ this phase 
exists, and whether its disappearance and 
the appearance
appearance
 of the usual hadronic phase (with
confinement and chiral  symmetry breaking)
is actually the same line, or some intermediate phase may also
exist in between.

F.Sannino (Yale)\footnote{He partially presenting his own talk and also
 substituted T.Appelquist who got ill right
before the talk} has started this discussion.
Based on the gap equation with the one-gluon exchange, Appelquist and
collaborators \cite{App}
have argued that it should happen close to the
line $N_f= 4 N_c$, or 12 flavors in SU(3). 
Another idea suggested by Appelquist et al is the so called ``thermodynamical
inequality", according to which the number of massless hadronic degrees 
of freedom $N(T=0)$ can never be larger than the number of fundamental
degrees of freedom $N(T=\infty)$.
The corresponding numbers at temperature T are defined as 
$$N= -F(T)*(90/\pi^2T^4) $$
 If the saturation of it, $N(0)=N(\infty)$, indicates
the boundary of hadronic world\footnote{Although I do not
understand the reasoning here, sorry. It may
somehow be related to 't Hooft matching anomaly conditions,
but I was not able to work it out.}, one can compare the number of pions
$N_\pi=(N_f^2-1)$ to the number of gluons and quarks (taken with the
coefficient 7/8) and get the same boundary as above. 

One may compare these ideas to the boundary found by Seiberg
based on his duality considerations and 't Hooft matching anomaly conditions. According to those,
the lower boundary of the conformal phase in N=1 SUSY QCD\footnote{Of course,
the ordinary and SUSY QCD have different multiplets and beta functions,
so we do not mean compare the numbers literally.} is at
$N_f=(3/2)N_c$. The ``thermodynamical
inequality" of Appelquist remarkably reproduces it!

However (as pointed out by Appelquist et al themselves)
he one gluon exchange gap equation actually indicate a $different$
point, and, even more important, a completely different pattern of
massless particles. The gap equation leads to quark and gluino chiral condensation,
but the
Seiberg phase  has a different set of massless hadrons
which are $not$ Goldstones,
related to chiral symmetry breaking. It probably mean that this approach
is too naive. Len me made a suggestion here: one can also get gap
equations for the channels favored by Seiberg and see if those can
make massless hadrons instead.

As we already mentioned in the section about finite T transition, the
instantons can restore chiral symmetry by breaking the 
random liquid into finite clusters, e.g. $\bar I I$ molecules.
With increasing $N_f$ this is also happens: it is easy to see
if one consider any fermionic line between them as a kind of additional
chemical binding bond. At some critical number of those, the entropy
of the random phase is no longer able to compensate for binding energy.
Explicit simulations suggest it to be at $N_f=5$,
above which the instanton-induced
chiral symmetry breaking disappears.
 This number agrees with a rapid change of the condensate value
 between $N_f=3$ and 4 (Mawhinney) and it is also
 much closer to lattice
indications (Iwasaki et al)  to the critical point  at $N_f=7$.
On the other hand, formation of instanton molecules by no means   prevents
 chiral symmetry breaking by a gluon exchange or any
other mechanism (confinement?), and so strictly speaking
 there is no direct contradiction between two approaches. 
One may have a strong decrease in a condensate, but not to zero at such $N_f=5-7$.

M.Velkovsky (BNL) discussed a calculation \cite{SVel} 
of the vacuum energy density
due to such  $\bar I I$ molecules. He concluded that for $N_f>6$ 
there is a difference between even and odd $N_f$: while for the former the
contribution vanishes, for the later it oscillate, changing the sign.
It may lead to different (or even alternating) phases  
at some intermediate $N_f$.

  A very interesting question  discussed by Sannino \cite{sannino}(see
  also \cite{Miranskii}) was 
a question about behavior near the conformal phase boundary.
He emphasized that the transition should be infinite order, with not
just few but $all$ hadronic masses going to zero (see also \cite{Miranskii}).

One particular pair of the correlators was discussed by Sannino in particular:
those are of two vector and axial correlators. In QCD they are related to
 rho  and a1 excitations, with their parameters approximately 
related to each other by two Weinberg sum rules\footnote{Those have zero r.h.s.
 because QCD have no operators of dimension 2, and also because
in the chiral limit the operator of the dimension 4, $G_{\mu\nu}^2$, cancels
in the difference. 
} should look
like. He has shown that as one becomes close to the transition in question, there
appear three separate momenta scales: (i) ``partonic" one, $p> \Lambda$,
(ii) ``hadronic" one $p< |<\bar q q >|^{1/3}$, and (iii) conformal window
in between. The 
contribution of the part (iii) to Weinberg sum rule, if non-zero,
 may deform the ``hadronic" theory compared to the usual QCD.

V.Elias (U. of Western Ontario)
 using Pade-summation for beta function, in SUSY and non-SUSy theories
E.Gardi (l'Ecole Polytechnique) to penetrate to the boundary of the
conformal window, and how far in $N_f/N_c$ can the perturbative theory
can actually be used. He concluded that for low $N_f$ such as zero
Pade approximant show no indications fro infrared fixed point. He also
discussed Kogan-Shifman scenario which appears due to a pole (rather
than zero) in the beta function.

E.Gardi also considered the boundary of the conformal window, both in
the ordinary and SUSY QCD. He emphasized that bottom of the window
correspond to $\gamma=0$. He concluded in particularly that QCD remains
weakly couple in the whole window, which excluded dual description. 
In SUSY QCD, on the other hand, does become strongly coupled inside
the window.

There was a discussion on how exactly people should look for this
transition on the lattice. 
As the transition itself is of ``infinite order"
 because the scale of chiral symmetry breaking
is going to the infrared,  it should look like rapid decrease
of the condensate, with unusual extrapolation to zero.
The demonstration of the ``conformal window" is much however
 more straightforward,
as it amounts to finding power-like correlators.
One more way 
to see it is 
to study scaling and construct lattice beta function:  it should vanish
in the conformal  window. In principle, it should converge to the
same
behavior in the infrared no matter what is the initial charge in the
lattice Lagrangian. In reality, the closer it is to the fixed point
the better.

\section{Some lessons from Supersymmetric Theories}
On the onset, let me emphasize one general point. SUSY theories are
not
a separate class of gauge theories, but rather a particular points
on the phase diagrams. One can always enlarge this theories breaking
the supersymmetry (e.g. consider the same fundamental fields but
different coupling constants). Therefore all features which are not
directly caused by SUSY should be true in general. Our general aim is  
to understand those general dynamical features, to the extent known
results in SUSY points can help.

M.Mattis (Los Alamos) had reviewed the status of the 
instanton calculus for the super-symmetric theories. For N=2 SUSY QCD
(``Seiberg-Witten theory'') it agrees with expansion of the elliptic
curves if $N_f< 2N_c$ but not for the case $N_f= 2N_c$. 

Let me inject here a discussion of
  the amusing similarity between QCD and (its relative) the N=2 SUSY QCD have been
recently demonstrated in \cite{RRS}. It is related to the issue of
already mentioned ``chiral scale" 1 GeV. In QCD it is phenomenologically
known that this scale is not only the upper bound of effective theory but also
the lower bound on parton model description. However, one cannot really see it
from the perturbative logs: 1 GeV is several times larger than their 
natural scale, $\Lambda_{QCD}\sim 200 MeV$. In the N=2 SUSY QCD the answer is known:
effective theory at small $a$ (known also as ``magnetic" formulation) is separated
from perturbative region of large $a$ by a singularity, at which monopoles
become massless and also the effective charge blows up. How it happens also  
follows from Seiberg-Witten solution, see Fig.2. Basically the perturbative
log becomes cancelled by instanton effects,  long before the charge blows
up due to ``Landau pole" at $p\sim \Lambda$. It happens ``suddenly" because
instanton terms have strong dependence on $a$: therefore perturbative
analysis seems good nearly till this point.

For comparison, in QCD we have calculated effective charge with the instanton correction, as
defined by Callan-Dashen-Gross expression. All we did was to
put into it the present-day knowledge of the
instanton density. The resulting curve is astonishingly similar to
the one-instanton one in N=2 SUSY QCD. Note, that  in this case as well,
 the ``suddenly
appearing" instanton effect blows up the charge, making perturbation
theory inapplicable, and producing massless pions, the QCD ``magnetic"
objects.  Moreover, it even happens at about the same place! (Which is
probably a coincidence.) 

\begin{figure}[t]
\vskip -0.4in
\epsfxsize=3.8in
\centerline{\epsffile{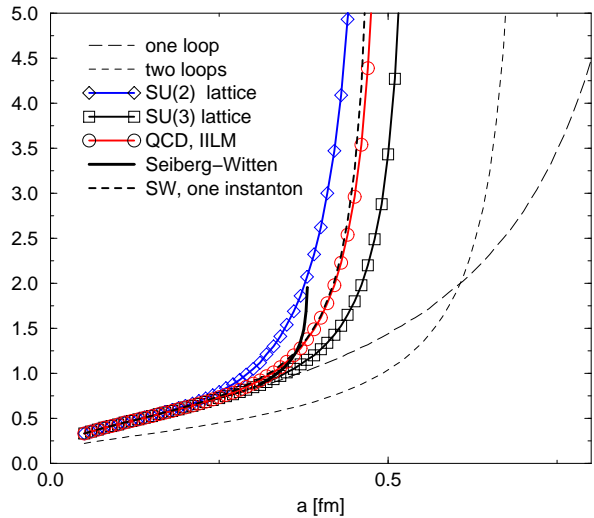}}
\vskip -0.05in
\caption[]{
 \label{figure1}
 The effective charge $b \,g^2_{eff}(\mu)/8\pi^2$ (b is the coefficient
of the one-loop beta function) versus normalization scale $\mu$ (in units of
its value at which the one-loop charge blows up). The thick solid line
correspond to exact solution \cite{SW_94} for the N=2 SUSY YM, the thick dashed line
shows the one-instanton correction. Lines with symbols (as indicated on figure)
stand for N=0 QCD-like theories,
SU(2) and SU(3) pure gauge ones and QCD itself. Thin long-dashed and short-dashed lines are one and two-loop results.
}
\end{figure}

The behavior is shown in  Fig.\ref{figure1}, where we
have included both a curve which shows the full 
coupling (thick solid line), as well as a curve which illustrates
only the one-instanton correction (thick dashed one). Because we will want
to compare the running of the coupling in different theories,
we have plotted $b g^2/8 \pi^2$ (b=4 in this case is the one-loop coefficient
of the beta function) and measure all quantities 
in  units of $\Lambda$, so that  the one-loop
charge blows out at 1. The meaning of the scale
can therefore be determined by what enters in the logarithm.

The title of Mattis 
 talk is actually ''The Physicist's proof of the Maldacena conjecture''.
In essence, this work \cite{Mattis_etal} is a semi-classical calculation of some
specific Green functions
in $N=4$ super-symmetric gauge theory\footnote{E.g. in N=4 theory considered
by Mattis all logs are gone and beta function is just zero.}, in the large number of
colors limit. The multi-instanton ``molecules" in this limit becomes dominated
by a configuration in which
all instantons are at the same place z and have the same size $\rho$:
there is enough space in color space not to worry about their overlap.
So, instanton is the ``master field" of this approach.
The answer obtained is in perfect agreement with Maldacena
conjecture and IIB SUGRA calculation, since it looks like
classical Green function in which all field propagate  from
the origination  points $x_1...x_n$
to a point in the $AdS_5$ space, which is nothing but\footnote{Let me recall
that when I found it, I had a feeling similar to the famous Mollier
character, who just discovered that in all his previous life what he was
saying and writing was ``prose". 
} $d^4zd\rho/\rho^5$.
Additional $S_5$ also appears, but as a non-trivial space of diquark
``condensates" created by such molecules.

\section{Topological effects in Applications}

  There were other workshops around (including two October RIVEN workshops
and November one in Nordita) dealing with QGP and the phase transition
as studied in heavy ion collisions. For that
reason we only included in our workshop those talks which have 
significant overlap with other discussions, such as 
 topology\footnote{Not directly related to instantons, which are
   discussed
in other sections.} and/or CP violating phases in the $\theta$ direction.
 
A.Zhitnitsky (Vancouver) had literally shocked the audience by his
bold proposal that the baryon asymmetry of the Universe
is $not$ due to baryon number violation but rather a
large scale {\it baryon charge separation} 
in the cosmological QCD transition \cite{Z}.
 He also proposed that all anti-quarks
are get locked in  the surface of
what he calls B-shell, now making the dark matter.
The reason it is locked is similar to domain wall fermions: it is a
topological
bound state resulting from different vacua inside and outside the
ball.
The sign of the charge is always the same, he explained, because the
vacuum inside has a particular CP phase. This meta-stable vacuum 
related to the (so far rather murky) subject of ``other brunches'' of
QCD
vacua as a function of $\theta$ parameter.
  
 This development is at its early stage, and it is not possible to
 tell
if it can survive. 
In a very lovely
discussion to follow, several critical comments were made. One of them
I made
are related to
safety issues related to fall
on by one of those shells. According to some estimates presented, the
baryon charge
of the ball is about $B\sim 10^{20}$, or a mass of the order of a gram.
If its energy is released in annihilation with matter, it is about an
atomic bomb. However Zhitnitsky argued that because 
the B-shells are large  bubbles of another vacuum,
the probability of the annihilation should be small.

M.Sadzikowski \cite{Sadzikowski}  (Cracow) has demonstrated that earlier estimates
of multiple production of baryons and anti-baryons in hadronic and nuclear collisions as a topological defects in chiral models was actually 
too optimistic. Including realistic quark masses and fluctuations
in the same model
significantly reduce the rate. His prediction for the rate is about
$10^{-4}$
anti-baryons/fm$^3$.

D.Kharzeev (BNL)  addressed
the issue of the non-trivial vacuum bubbles with effectively different
$\theta$ and CP violation \cite{PK}. Unlike Zhitnitsky, however, he discussed
 heavy ion collisions, not cosmology. He argued that high-degree
of U(1) restoration may make it possible, although in small vicinity
of $T_c$.
The  estimates of what
the probability of such bubble production  are very uncertain. However some  
ideas how one should look for it were
discussed.
\section{Acknowledgements}  The organizers of the workshop
 are grateful to  RIKEN/BNL Center
for its support, and to all the speakers for their inspiring work.
 
\end{document}